# A biomechanical model of swallowing for understanding the influence of saliva and food bolus viscosity on flavour release

Clément de Loubens[a,b,*], Albert Magnin[c], Marion Doyennette[a,b],

Ioan Cristian Tréléa[b,a], Isabelle Souchon[a,b]

[a] INRA, UMR 782 Génie et Microbiologie des Procédés Alimentaires, CBAI 78850 Thiverval Grignon, France

[b] AgroParisTech, UMR 782 Génie et Microbiologie des Procédés Alimentaires, CBAI 78850 Thiverval Grignon, France

[c] Laboratoire de Rhéologie, Université Joseph Fourier-Grenoble I, Grenoble INP, CNRS (UMR 5520), BP 53, Domaine Universitaire, 38041 GRENOBLE cedex 9, France

∗Corresponding author

Email addresses: cdeloubens@grignon.inra.fr (Clément de Loubens), magnin@ujf-grenoble.fr (Albert Magnin), souchon@grignon.inra.fr (Isabelle Souchon)

**Abstract**

After swallowing a liquid or a semi-liquid food product, a thin film responsible for the dynamic profile of aroma release coats the pharyngeal mucosa. The objective of the present article was to understand and quantify physical mechanisms explaining pharyngeal mucosa coating. An elastohydrodynamic model of swallowing was developed for Newtonian liquids that focused on the most occluded region of the pharyngeal peristaltic wave. The model took lubrication by a saliva film and mucosa deformability into account. Food bolus flow rate and generated load were predicted as functions of three dimensionless variables: the dimensionless saliva flow rate, the viscosity ratio between saliva and the food bolus, and the elasticity number. Considering physiological conditions, the results were applied to predict aroma release kinetics.

Two sets of conditions were distinguished. The first one was obtained when the saliva film is thin, in which case food bolus viscosity has a strong impact on mucosa coating and on flavour release. More importantly, we demonstrated the existence of a second set of conditions. It was obtained when the saliva film is thick and the food bolus coating the mucosa is very diluted by saliva during the swallowing process and the impact of its viscosity on flavour release is weak. This last phenomenon explains physically *in vivo* observations for Newtonian food products found in the literature. Moreover, in this case, the predicted thickness of the mix of food bolus with saliva coating the mucosa is approximately of 20 μm; value in agreement with orders of magnitude found in the literature.

Keywords: lubrication, pharynx, elastohydrodynamic, viscosity, aroma

# 1. Introduction

Food formulation has to take different recommendations to improve nutritional quality of foods (low fat content, less salt and sugar) and to adapt food to specific people (as disphagic patients) without modifying their organoleptic qualities (flavour and texture perception). These organoleptic qualities are closely related to the physiological process of food transformation during chewing and swallowing (Weel et al., 2004; Boland et al., 2006). It is so necessary to study the processes of food bolus formation (Woda et al., 2010 ; Yven et al., 2010) and of swallowing mechanisms in relation with physical properties of food (Taniguchi et al., 2008; Tsukada et al., 2009) to formulate novel food products.

Swallowing of a liquid or a semi-liquid food product generates a thin film of product coating the pharyngeal mucosa (Levine, 1989) responsible for the dynamic profile of aroma release (Buettner et al., 2001). The influence of rheology of liquid and semi-liquid food products on aroma release and perception is an unclear and debatable issue in the literature (Hollowood et al., 2002; Cook et al., 2003; Weel et al., 2004; Saint-Eve et al., 2006). We can assume that the conclusions did not match because the experimental investigations covered very different rheological properties (from yield stress fluids as yoghurt to shear-thinning fluids as hydrocolloids). Moreover, these analyses may have been biased by the fact that rheological properties and physico-chemical properties governing aroma relase (such as mass transfer coefficient, Tréléa et al., 2008) are often coupled properties of the product. To explain the role of product rheology on aroma release, we need to study the physical phenomena governing pharyngeal mucosa coating.

To understand these phenomena, de Loubens et al. (2010) analysed the physiology and biomechanics of swallowing. They showed that the thin film of product coating the mucosa is due to a weak reflux during the pharyngeal peristalsis between the root of the tongue and the posterior pharyngeal wall (Figure 1a). To physically represent this phenomenon and simplify the problem, they focused their attention on the most occluded region of the peristaltic wave. In this region, the pharyngeal peristalsis wave is equivalent to a forward roll coating process. Based on this physiological analysis, a fluid-mechanical model that considers lubrication by a saliva film was developed. However, mucosa deformability was not considered in the first model, whereas it is an important phenomena that may quantitatively improve the model predictions. In the present study, we consider that the pharyngeal peristalsis is equivalent to a forward roll coating process with deformable and lubricated surfaces (Figure 1b). In this process, the mucosa deform under the load L' applied by the pharyngeal constrictors muscles (Figure 1b). The purpose of this study was to develop an elastohydrodynamic model of the pharyngeal peristalsis in order to understand and quantify the role of saliva and the food bolus on the pharyngeal mucosa coating. The equation system was scaled by the elastic effects and solved numerically. A parametric study showed the influence of the different model parameters on food bolus flow rates and generated forces. The model was applied to flavour release and the predictions were compared with *in vivo* observations obtained for Newtonian liquid foods from the literature. Finally, main model assumptions were discussed.

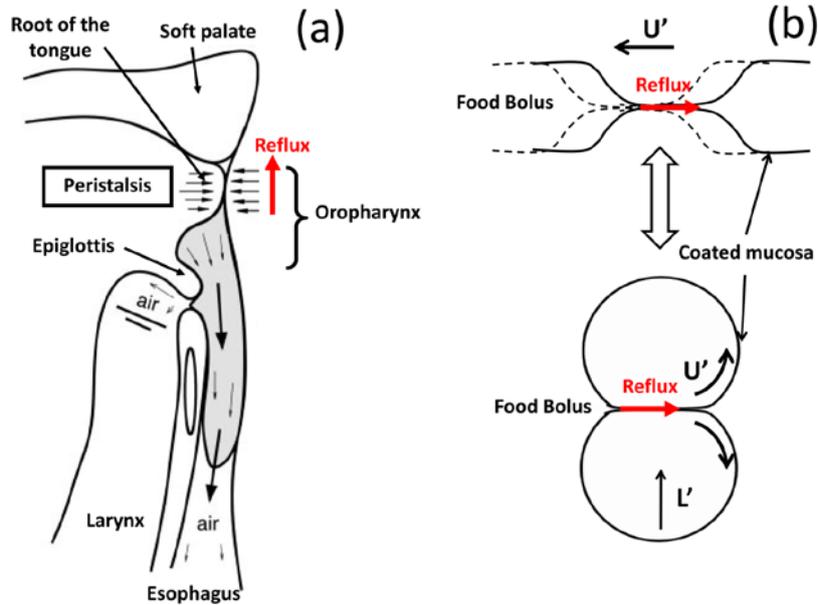

**Figure 1:** (a) Pharyngeal peristalsis (adapted from Pal et al., 2003). (b) Diagram of the peristaltic wave and associated study system. Near the most occluded point, the pharyngeal walls are in rotation compared to each other. U' is the wave velocity (m/s) and L' the load applied by the pharyngeal constrictors muscles (N/m), adapted from de Loubens et al. (2010).

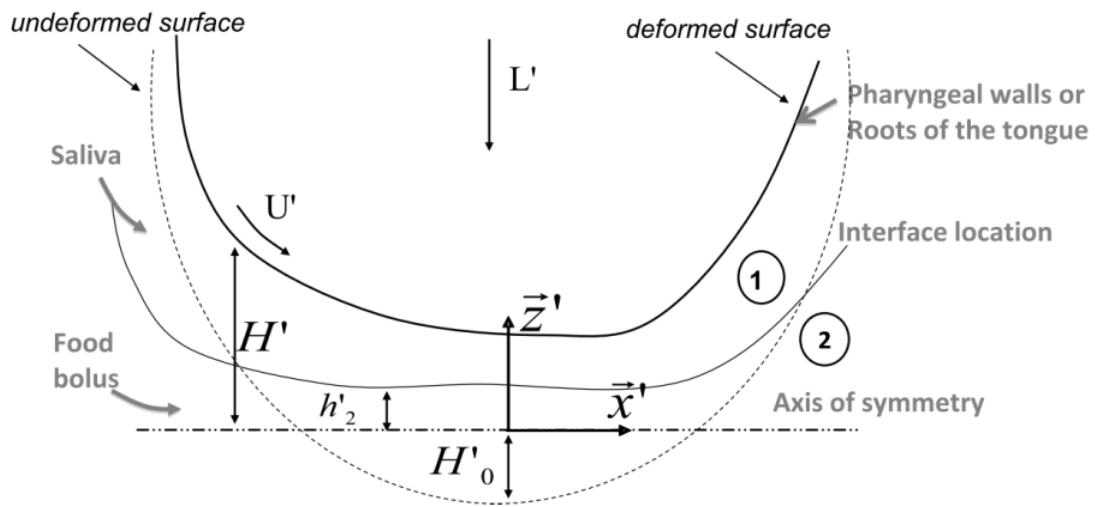

**Figure 2:** Diagram of definition and notations. U' is the wave velocity (m/s), L' the load applied by the pharyngeal constrictors muscles (N/m), H'(x') the mucosa location (m), $h_2'$ (x') the interface location between the food bolus and the saliva and $H_0'$ the negative-gap width.

## 2. Elastohydrodynamic model of the pharyngeal peristalsis

*2.1. Model hypothesis*

As de Loubens et al. (2010), we considered that the swallowing process is equivalent to a forward roll coating process (Figure 1). Moreover, we took the deformability of the mucosa into account. The general features of the forward roll coating process with deformable rolls for Newtonian fluids have been described by Coyle (1988). This author analysed the flow by means of two dimensionless numbers: the elasticity parameter $E_s$ is the ratio of viscous to elastic forces:

$$E_s = \frac{\mu_0' U'}{(2R')^2 E_e'} \quad (1)$$

and the load parameter F is the ratio of the external load to the elastic forces:

$$F = \frac{L'}{(2R')^2 E_e'} \quad (2)$$

where $E_e'$ is the effective elastic modulus of the substrate that covers the deformable rolls (Pa/m), μ' the viscosity of the fluid (Pa.s), R' the rolls radius (m), U' the velocity (m/s) and L' the applied load per unit of width (N/m).

**Table 1:** Physiological variables and approximate corresponding values.

| Description | Symbol | Typical values | References |
|---|---|---|---|
| Saliva thickness | $e_1'$ | no data | |
| Saliva viscosity | $\mu_1'$ | 1 1-10 mPa.s | Schipper et al., 2007; Stokes et al., 2007 |
| Bolus viscosity | $\mu_2'$ | > 1 mPa.s | |
| Wave velocity | U' | 0.1-0.5 m/s | Dantas et al., 1990; Meng et al., 2005; Chang et al., 1998 |
| Radius | R' | 40 mm | estimated from Chang et al., 1998 |
| Elasticity modulus of the mucosa | E' | 20-200 kPa | Diridollou et al., 2000; Kim et al., 1998 |
| Mucosa thickness | $e_m'$ | 1-4 mm | Diridollou et al., 2000 |
| Load | L' | 10-60 N/m | de Loubens et al., 2010 |
| Elasticity parameter | $E_s$ | ~ $8.10^{-9}$ | calculated with (1) |
| Load parameter | F | ~ $3.10^{-5}$ | calculated with (2) |

Two limiting cases can be distinguished (Johnson, 1970). When F is low and $E_s$ is high, the viscous forces predominate. This case tends to the rigid roll limit that was the case developed for pharyngeal peristalsis by de Loubens et al. (2010). When F is high and $E_s$ is low, the elastic forces dominate and the pressure profile is similar to that of a dry contact. This case is the large deflection limit. The cylinders surfaces would intersect if there were no deformation. Coyle (1988) defined the effective elastic modulus by $E_e' = E'/e_m'$, where E' is the Young modulus of the substrate (Pa) and $e_m'$ its thickness (m). Useful physiological data on the pharyngeal peristalsis are given in Table 1. From these data and the results obtained by Coyle (1988), we can estimate that the pharyngeal peristalsis occurs on the large deflection limit (F ≈ $3.10^{-5}$ and $E_s$ ≈ $8.10^{-9}$), although the parameters have a wide range of variation.

The present physical situation is therefore modeled with the lubrication approximation: the inertial terms are neglected compared to the viscous terms in the Navier-Stokes equations. The use of the lubrication approximation for the most occluded region of the pharyngeal peristalsis wave and the fact that the flow can be considered as stationary was already justified by de Loubens et al. (2010). In addition, we take the presence of a lubricating saliva film and mucosa deformability into consideration.

Since the confusion concerning the role of food rheology on flavour release, we restrict our analysis to homogeneous Newtonian food bolus. Moreover, in the paragraph concerning the model applications (4.2), model predictions were compared with *in vivo* data obtained with Newtonian glucose solutions. As demonstrated by de Loubens et al. (2010), the main role of saliva during swallowing is to obstruct the contact. To represent this phenomenon, saliva is considered as being a Newtonian fluid too.

The geometry is symmetric along the x-axis (Figure 2). Relative quantities associated with saliva and the food bolus are referred to as 1 and 2, respectively. Between the two fluids, we ignored diffusion and surface tension effects. The dimensional values are identified by the symbol '. The flow rate of saliva $q_1'$ ($m^3/s$) is assumed to be known and the flow rate of the food bolus $q_2'$ is calculated. $\mu_i'$ (Pa.s) refers to the viscosities, $e_m'$ the thickness (m) of the deformable layer of mucosa, $H'(x)$ the half gap between the two cylinders (m), $H_0'$ the "negative-gap width" (m), $h_2'(x)$ the location of the interface between the food bolus and saliva (m), $U'$ the cylinder velocity (m/s), $L'$ the load per unit of width (N/m), and $R'$ the radius (m).

*2.2. Elastic model of the mucosa*

Near the contact point, the undeformed roll surface profiles are locally approximated by parabolas:

$$H'(x') = -H_0 + \frac{x'^2}{2R'} + \Delta H'(x') \quad (3)$$

where $\Delta H'(x')$ is the cylinder surface deflection and must be expressed in terms of model for the elastic deformation of the rolls. The deformation of the layer can be considered with different models. Skotheim and Mahadevan (2005) have carried out a detailed study of fluid-immersed compressible, incompressible and poro-elastic soft interfaces. The one-dimensional Constrained Column Model (CCM) is the most tractable and the least intensive at the computational level. It assumes that the local pressure $p'$ is directly proportional to the local deflection $\Delta H'$:

$$\Delta H'(x') = \frac{p'(x')}{E_e} \quad (4)$$

For large deflections and incompressible compliant layers such as mucosa, Carvalho and Scriven (1995) and Gostling et al. (2003) have proposed:

$$E_e' = \frac{4E'}{e_m'} \quad (5)$$

They found good agreement between this model and most of the sophisticated models in terms of the flow rates and the generated forces. These two last assumptions were retained to model the surface deflection (Eq. 4 and 5).

*2.3. Dimensionless variables*

For high load, viscous forces are small compared to elastic forces, so the pressure should be scaled with the latter. Choosing $H'_0$ as the length scale is the most convenient choice because it allows the model to be written in two parameters only, namely the viscosity ratio:

$$\alpha = \frac{\mu'_2}{\mu'_1} \quad (6)$$

and the elasticity number:

$$N_e = E_S \left(\frac{2R'}{H'_0}\right)^{5/2} \quad (7)$$

where $E_s$ is defined with the saliva viscosity: $E_S = \frac{\mu'_1 U'}{(2R')^2 E'_e}$.

The limit Ne→+∞ corresponds to the case where the undeformed rolls would touch. The limit Ne→0 corresponds to the dry rolling contact. The dimensionless values defined for imposed velocity and gap are given by:

$$x = \frac{x'}{\sqrt{2R'H'_0}}$$

$$z = \frac{z'}{H'_0}$$

$$u_i = \frac{u'_i}{U'}$$

$$q_i = \frac{q'_i}{U'H'_0}$$

$$p_i = \frac{p'_i}{E'_e H'_0}$$

$$L = \frac{L'}{E'_e H'_0 \sqrt{2R'H'_0}}$$

*2.4. Hydrodynamic model*

The cylinder profile is given by:

$$H(x) = -1 + x^2 + p(x) \quad (8)$$

The momentum conservation equations are solved in the lubrication approximation in their dimensionless form:

$$\frac{\partial p}{\partial x} = N_e \frac{\partial^2 u_1}{\partial z^2} \quad (9)$$

$$\frac{\partial p}{\partial x} = \alpha . N_e \frac{\partial^2 u_2}{\partial z^2} \quad (10)$$

$$\frac{\partial p}{\partial z} = 0 \quad (11)$$

Defining η = z/H(x) and β = $h_2$(x)/H(x), and considering no wall slip, continuity of velocity and shear stress at the interface between the food bolus and the saliva and symmetry, the boundary conditions are:

$$u_1(\eta = 1) = 1 \quad (12)$$

$$u_1(\eta = \beta) = u_2(\eta = \beta) \quad (13)$$

$$\left.\frac{\partial u_1}{\partial \eta}\right|_{\eta=\beta} = \alpha \left.\frac{\partial u_2}{\partial \eta}\right|_{\eta=\beta} \quad (14)$$

$$\left.\frac{\partial u_2}{\partial \eta}\right|_{\eta=0} = 0 \quad (15)$$

After integration of (9) and (10), application of the boundary conditions (12), (13), (14) and (15) and of the mass conservation, the flow rates are given by:

$$q_1 = \frac{H^3(\theta)}{2N_e}\cos^2(\theta)\frac{dp}{d\theta}\left[-\frac{\beta^3}{3} + \beta - \frac{2}{3}\right] + H(\theta)(1-\beta) \quad (16)$$

$$q_2 = \frac{H^3(\theta)}{2N_e}\cos^2(\theta)\frac{dp}{d\theta}\left[\beta^3\left(1 - \frac{2}{3\alpha}\right) - \beta\right] + H(\theta)\beta \quad (17)$$

where θ = arctan(x) .

Upstream, we consider that the contact is fully submerged. Downstream, the film splits. In the large deflection case, Coyle (1988) has demonstrated that this boundary condition has a slight effect on the results, so we consider that:

$$p\left(-\frac{\pi}{2}\right) = p\left(\frac{\pi}{2}\right) = 0 \quad (18)$$

After resolution, we calculate the resulting load:

$$L = \int_{-\frac{\pi}{2}}^{\frac{\pi}{2}} \frac{p(\theta)}{\cos^2(\theta)} d\theta \quad (19)$$

*2.5. Resolution method*

From (16) and (17), we obtain an algebraic equation:

$$\frac{2}{3}H(\theta)\left(1-\frac{1}{\alpha}\right)\beta^4 + \left[(q_1 - H(\theta))\times\left(1-\frac{2}{3\alpha}\right)+\frac{q_2}{3}\right]\beta^3 + \left[\frac{H}{3}-q_1-q_2\right]\beta + \frac{2}{3}q_2 = 0 \quad (20)$$

and a differential equation on the pressure:

$$\frac{dp}{d\theta} = \frac{3Ne(q_1 + q_2 - H(\theta))}{H(\theta)^3 \cos(\theta)^2 \left[\beta^3\left(1-\frac{1}{\alpha}\right)-1\right]} \quad (21)$$

Where

$$H(\theta) = -1 + \tan(\theta)^2 + p(\theta) \quad (22)$$

Equations (20), (21) and (22) were solved using Matlab7 software. Even so, the integration had to be performed backwards in space (from π/2 to -π/2) to obtain numerical stability. $q_2$ is the unknown variable. For a set of parameters ($q_1$, α, Ne), we iterated on $q_2$ until the boundary conditions (18) were verified.

## 3. Parametric study

*3.1. Mono-layer case*

Numerical solutions were validated by comparing the results in the monolayer case with those of Coyle (1988). Figure 3 shows the flow rate $q_1$ and the load L as a function of the elasticity number $N_e$. As shown by Coyle (1988), from the results presented Figure 3, the flow rate and load dependence with $N_e$ can be approximated by the relationships:

$$q_1 \approx 0.5 N_e^{0.5} \text{ when } q_2 = 0 \quad (23)$$

$$L \approx 1.3 + 1.7 N_e^{0.55} \text{ when } q_2 = 0 \quad (24)$$

When $N_e$ tends to zero, the flow rate decreases and the load tends to 1.3. This value corresponds to a dry rolling contact and was verified analytically (Coyle, 1988).

*3.2. Food bolus flow rates*

Figure 4-a shows the influence of Ne and Figure 4-b the influence of the saliva flow rates $q_1$ at $N_e=1$ on the food bolus flow rates $q_2$ for different cases. The food bolus flow rate $q_2$ decreases when $N_e$ tends to zero corresponding to the dry rolling contact. When there is no saliva at the interface ($q_1=0$), $q_2$ dependence with Ne and α can be expressed with a relationship similar to (23):

$$q_2 \approx 0.5(\alpha N_e)^{0.5} \text{ when } q_1 = 0 \quad (25)$$

Increasing the viscosity ratio α increases $q_2$ whereas saliva lubrication decreases $q_2$. The influence of the saliva flow rate $q_1$ decreases when $N_e$ increases. When the relationship:

$$N_e \approx 4q_1^2 \quad (26)$$

is verified, the contact is over-flooded by saliva and $q_2$ tends to zero.

The viscosity ratio α has a strong influence on the food bolus flow rate $q_2$ when the saliva flow rate $q_1$ is low. Its impact drop sharply when $q_1$ increases.

*3.3. Load*

Figure 5-a shows the influence of $N_e$ and Figure 5-b the influence of $q_1$ at Ne=1 on the generated load L for different cases. When the contact is not lubricated by saliva, we obtain a relationship equivalent to (24):

$$L \approx 1.3 + 1.7(\alpha N_e)^{0.55} \text{ when } q_1 = 0 \quad (27)$$

When α increases, L increases. When α is smaller than 1, L decreases with $q_1$, whereas L increases with $q_1$ when α is higher than 1. The dependence of L on $q_1$ is highly reduced when $N_e$ is weak due to the fact that the contribution of hydrodynamic pressure to the load is negligible.

*3.4. Pressure profile*

Figures 6-a and b show pressure profiles for $N_e=1$ and $N_e=10^{-3}$, respectively, for different cases. The pressure sharply increases as the fluid is dragged into the narrowing channel, after which the channel widens and the pressure drops. When α is higher than 1 the pressure profile developed with α and the saliva flow rate $q_1$ reduces its development and, inversely, when α is lower than 1. When Ne is weak, the pressure profile is less dependent on α and $q_1$ as shown in Figure 6-b for Ne = $10^{-3}$. It tends to a parabola corresponding to a dry rolling contact (Coyle, 1988): the pressure profile is dominated by the elastic deformation of the mucosa.

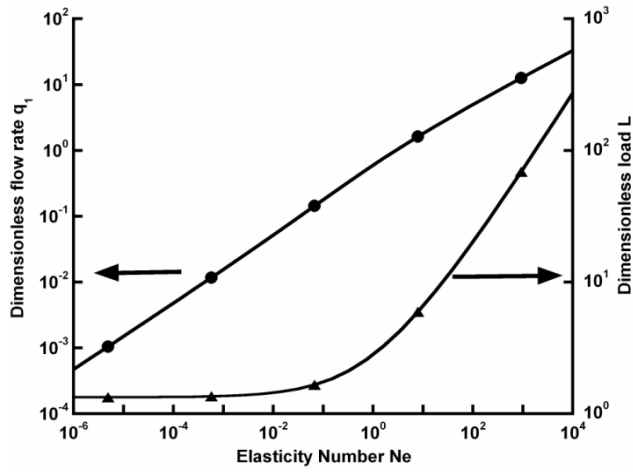

**Figure 3:** Dimensionless flow rate $q_1$ (-●-) and load L (-▲-) as a function of the elasticity number $N_e$ in the mono-layer case ($q_2=0$).

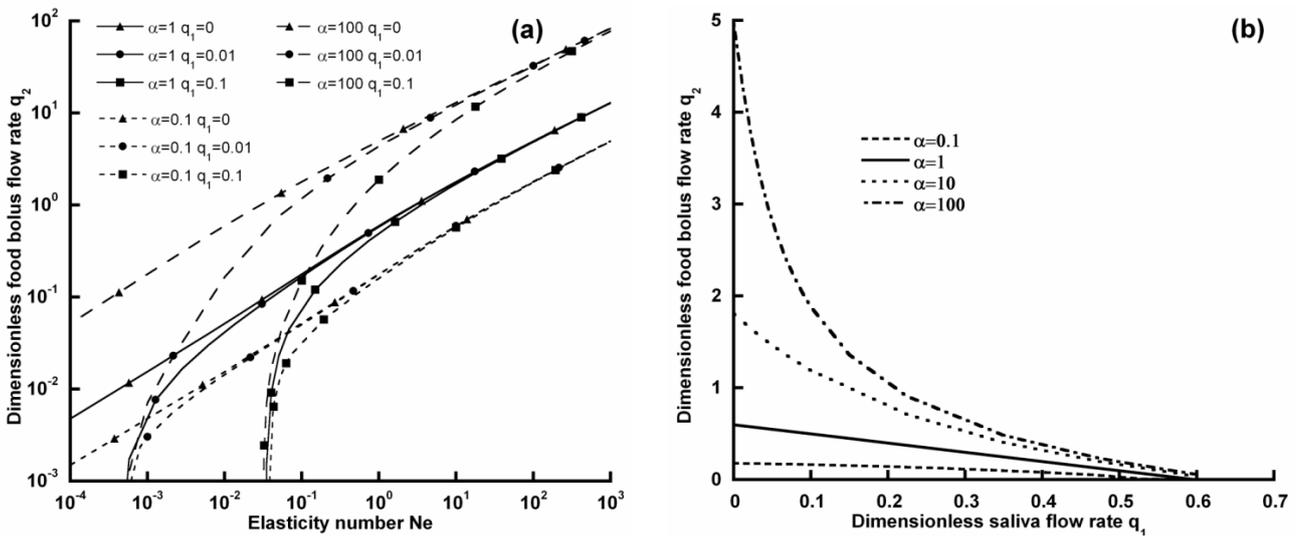

**Figure 4:** Dimensionless food bolus flow rate $q_2$ as a function of the elasticity number $N_e$ for different viscosity ratios α and dimensionless saliva flow rates $q_1$ (a) and as a function of dimensionless saliva flow rate $q_1$ for different viscosity ratios α for $N_e=1$ (b).

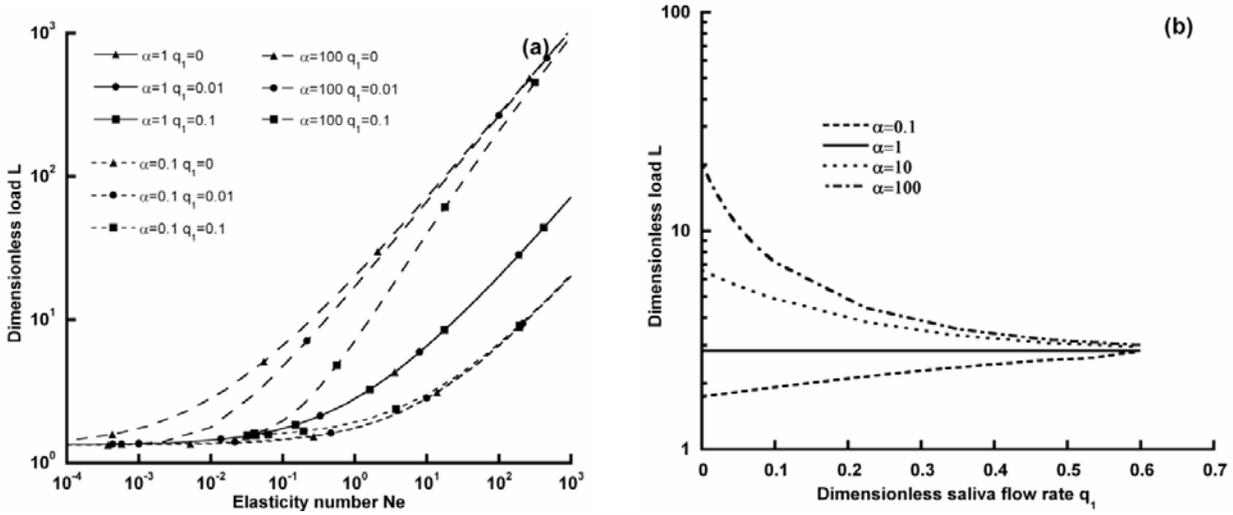

**Figure 5:** Dimensionless load L as a function of the elasticity number $N_e$ for different viscosity ratio α and dimensionless saliva flow rates $q_1$ (a) and as a function of the dimensionless saliva flow rate for different viscosity ratios α at imposed gap and velocity for $N_e=1$ (b).

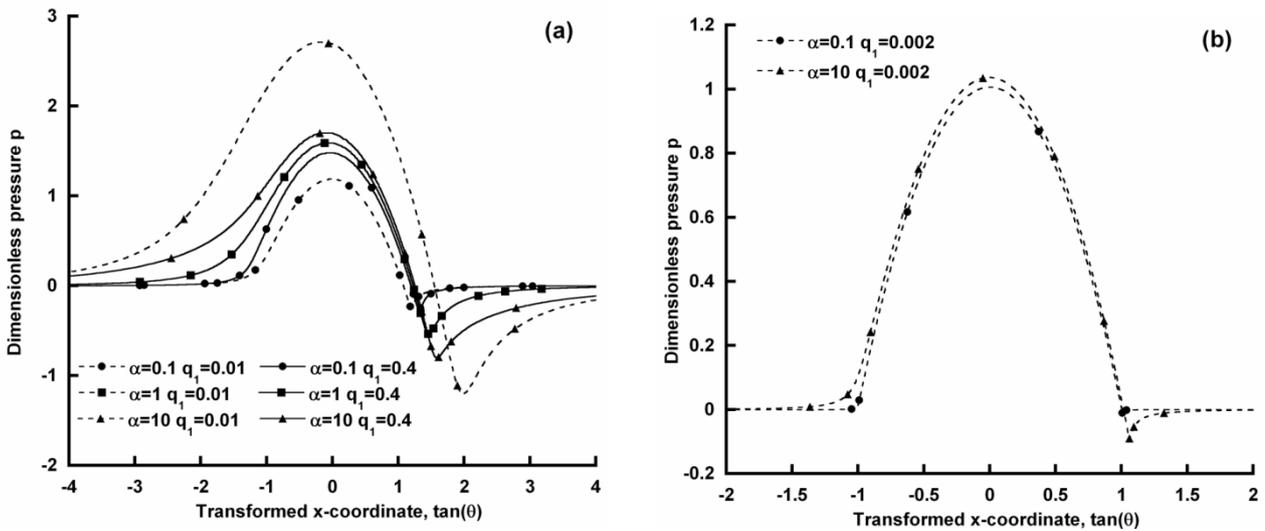

**Figure 6:** Dimensionless pressure profiles p for different viscosity ratio α and different dimensionless saliva flow rates $q_1$ at $N_e=1$ (a) and $N_e=10^{-3}$ (b).

**4. Applications**

The aim of this section is to provide quantitative results for typical physiological parameters, to apply these results to *in vivo* aroma release and to compare the predictions with *in vivo* experiments found in the literature.

*4.1. Application to swallowing*

Coating flows often present instabilities and the film varies in a wavy, sinusoidal-like manner across the substrate. This type of film thickness non-uniformity is usually referred to as ribbing. It is a consequence of an imbalance between surface tension forces and the pressure gradient present within the downstream nip region that generate vortex in the film-split region. In the case of a bi-layer coating, the two fluids are mixed together under the vortex action at the contact output. Chong et al. (2007) observed that ribbing is present over a wide range of operating parameters for negative gaps. We can thus consider that ribbing and vortex occur during swallowing and that the food bolus is therefore mixed with the saliva film. The interesting model outputs in terms of flavour release are the total thickness e' of the mixture of the food bolus with saliva (e'=e'$_1$+e'$_2$) and the rate of dilution r of the food bolus in saliva defined by:

$$r = 100 \frac{e_2'}{e_1' + e_1'} \quad (28)$$

In order to apply the model to pharyngeal peristalsis, the mathematical model was used to calculate the thickness of bolus e'$_2$ deposited on the pharyngeal mucosa at imposed velocity U' and load L'. A value of L' to be reached was fixed and (20), (21) and (22) were solved as explained in 2.5. We iterated on q'$_2$ and H'$_0$ until (18) and (19) were verified. In fact, the action of the pharyngeal constrictors muscles is equivalent to setting a normal force on the rolls, refferd to as load L' (de Loubens et al., 2010).

Figure 7 shows the total thickness (in μm) as a function of the rate of dilution (in %) for different parameters representative of different physiological conditions (Table 1).

Regardless of the parameters, the values of the elasticity number N$_e$ are lower than the 1 and, as previously explained, the situation is therefore similar to the dry rolling contact. The load is due to the elastic forces and not to the hydrodynamic pressure.

When the viscosity ratio α is 1 (cases a1, b1, c1), the deposited thickness is constant regardless of the dilution rate is. When the viscosity ratio increases (comparison between the cases a1 and a10, for example), there are two sets of conditions. The first one is obtained when the food bolus is not very diluted with saliva (r → 0%) and the viscosity ratio has a considerable influence on the total thickness e'. The second one is obtained when the food bolus dilution increases (r → 100%) and the total thickness tends to a constant.

In the cases a1, the rate of dilution between the two sets of conditions is about 45%, resulting in an initial saliva thickness e'$_1$ of approximately 5 μm.

When the dilution ratio is maximal, the saliva entirely obstructs the contact and the bolus cannot coat the mucosa. The limit value of saliva thickness is approximately 10 μm in the case a.

The comparison between cases a and b illustrates the strong role of the peristalsis wave velocity U' When U' is multiplied by 5, the total thickness is multiplied by 2.5. Moreover, the limit rate of dilution r and the limit of saliva thickness between the two sets of conditions previously described increase when the wave velocity increases: in case a they are about 45% and 5 μm and 55% and 15 μm in case b. The saliva thickness value necessary to over-flood the contact increases from about 10 to 25 μm (r = 100%) as well.

The comparison of cases a10 and c10 shows that increasing the Young modulus of the mucosa E' reduces the total thickness. The values of E' reported in Table 1 have one decade of dffer ence. This parameter is difficult to obtain *in vivo* and we have therefore used the Young modulus obtained from human skin *in vivo* (Diridollou et al., 2000) and of human pharyngeal tissue in *post mortem* tension (Kim et al., 1998). The mechanical behavior of the mucosa would require more considerations. In fact, mucosa presents a viscoelastic behavior (Kim et al., 1998) and, as a result, the Young modulus obtained at the time scale of the process should be introduced into the model (Cohu and Magnin, 1997).

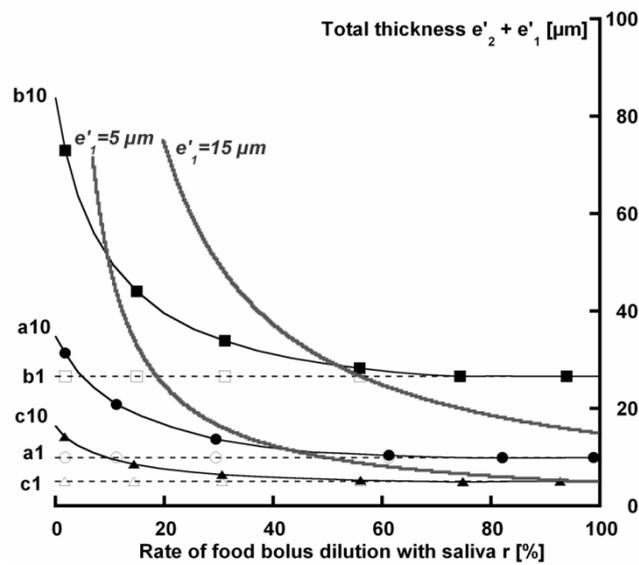

| Cas n° | U' [m/s] | E' [kPa] | $\mu_2'$ [mPa.s] | α | $E_s$ | F |
|---|---|---|---|---|---|---|
| a1 (○) | 0.1 | 20 | 5 | 1 | $4\ 10^{-9}$ | $8\ 10^{-5}$ |
| a10 (●) | 0.1 | 20 | 50 | 10 | $4\ 10^{-9}$ | $8\ 10^{-5}$ |
| b1 (□) | 0.5 | 20 | 5 | 1 | $2\ 10^{-8}$ | $8\ 10^{-5}$ |
| b10 (■) | 0.5 | 20 | 50 | 10 | $2\ 10^{-8}$ | $8\ 10^{-5}$ |
| c1 (△) | 0.1 | 200 | 5 | 1 | $4\ 10^{-10}$ | $8\ 10^{-6}$ |
| c10 (▲) | 0.1 | 200 | 50 | 10 | $4\ 10^{-10}$ | $8\ 10^{-6}$ |

**Figure 7:** Total thickness of food bolus and saliva $e'=e_1'+e_2'$ coating the pharyngeal mucosa as a function of the dilution rate of the food bolus with saliva $r=100\dfrac{e_2'}{e_1'+e_1'}$ and iso-values of saliva thickness $e_1'$ (grey lines) ($e_m'$=4 mm, R'=4 mm, $\mu_1'$=5 mPa.s, L'=10 N/m).

*4.2. Application to flavour release*

Predictions of aroma release kinetics

The results of the pharyngeal mucosa coating model were used in a mechanistic model that predicts aroma release (Doyennette et al., 2011). Figure 8 shows the kinetics of aroma release in the nasal cavity predicted by the mechanistic model for different viscosity ratio α and rates of dilution r calculated with the present model. In this section, we considered that the physico-chemical properties of the food bolus are independent of its viscosity.

Two sets of conditions can be distinguished according to the physiological parameters and the viscosity ratio. When the initial thickness of saliva and the dilution are weak (r→0%, cases 3 and 4), viscosity has a considerable effect on the decreasing part of the aroma release kinetics, whereas when the dilution with saliva is strong (r→100%, cases 1 and 2), viscosity has no effect on aroma release. Figure 7, we show that for typical physiological parameters and a food bolus viscosity of 50 mPa.s, the order of magnitude of the limit value of saliva thickness that distinguishes the two cases is between 5 and 15 μm.

Comparison with in vivo aroma release kinetics

In this section, the model predictions are compared with the results obtained in the literature.

Doyennette et al. (2011) carried out an *in vivo* investigation of the influence of viscosity on aroma release. They used glucose solutions as test fluids that varied widely in viscosity (from 0.7 to 405 mPa.s at 35°C). They concluded that the solution coating the pharyngeal mucosa was highly diluted with saliva. To show this, they compared the maximal relative concentration of kinetics $C_{max}$ obtained *in vivo* with their model predictions for two different cases.

Figure 9 shows the maximal concentration of kinetics $C_{max}$ obtained *in vivo* and predicted by the model in two different cases as a function of the viscosity of the glucose solution. They observed a maximal difference of 40% *in vivo* on $C_{max}$, depending on the glucose viscosity of the solution. However, when they simulated aroma release kinetics by considering that the residual thickness of the product was not diluted by saliva (r=0%), they observed differences of 97% between the products whereas, when they considered a rate of dilution r of approximately 85%, their predictions were in agreement with the *in vivo* observations. Thus, it was necessary to suppose that the food bolus was highly diluted by saliva to explain the *in vivo* observations.

The biomechanical model developed in the present study makes it possible to understand the physical origins of these observations: the initial thickness of saliva coating the mucosa is sufficiently thick to dilute the food bolus coating the mucosa at the level of the most occluded region of the pharyngeal peristaltic wave and to break the viscosity influence on coating and flavour release. Moreover, the thickness of the residual film that coats the mucosa after swallowing was estimated at approximately 15 μm in their study and this value is close to those calculated with the present model (Figure 7).

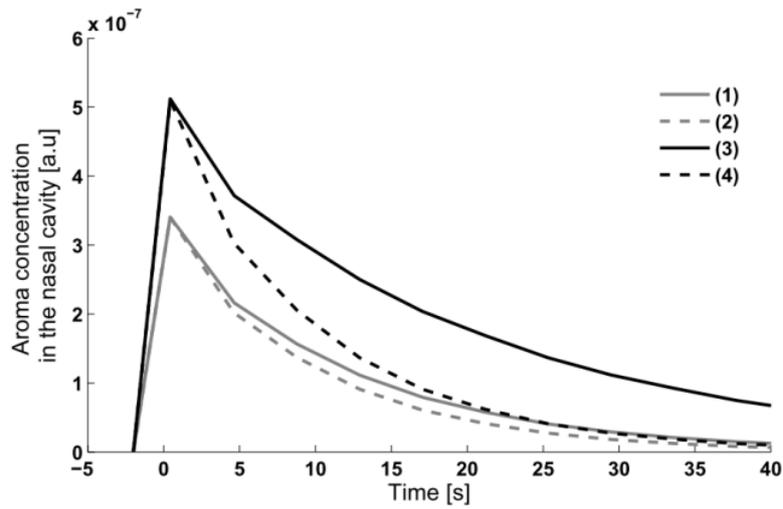

| Cas n° | α | r (%) | e'₂+e'₁ |
|---|---|---|---|
| 1 | 10 | 40 | 12 |
| 2 | 1 | 40 | 10 |
| 3 | 10 | 10 | 20 |
| 4 | 1 | 10 | 10 |

**Figure 8:** Aroma release kinetics predicted by the mechanistic model developed by Doyennette et al. (2011) for different rates of dilution of the food bolus with saliva ($r = 100\frac{e_2'}{e_1'+e_1'}$) and total thicknesses ($e_1'+e_2'$) predicted with the present elastohydrodynamic model. The time 0 s corresponds to the swallowing events. (U'=0.5 m/s, E'=20 kP a, $e_m$'=4 mm, R'=4 mm, $\mu_1$'=5 mPa.s, L'=10 N/m )

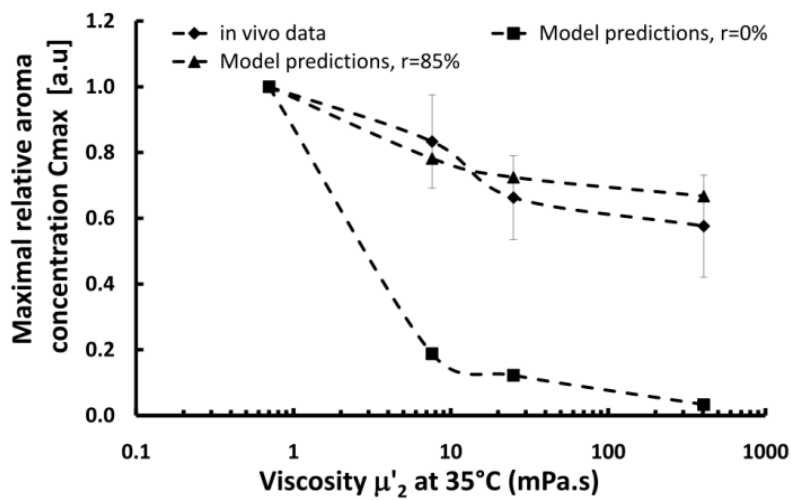

**Figure 9:** Maximal relative concentration of aroma release kinetics $C_{max}$ as a function of the viscosity of glucose solutions $\mu_2$': *in vivo* data (♦), model predictions without dilution with saliva (r=0%,■), model predictions with a rate of dilution of product with saliva r of 85% (▲). Error bars represent the standard deviation on the in vivo data. Data from Doyennette et al. (2011).

## 5. Discussion about non-Newtonian behavior

In despite of different assumptions performed in the model, this last is able to explain the physical origins of *in vivo* observations for Newtonian fluids. The main assumptions concern the physical fluids properties (saliva and food bolus) and especially their rheological behavior that we discuss in this section.

*5.1. Rheology of saliva*

In the present model of pharyngeal peristalsis, saliva was considered as a Newtonian fluid although it presents complex rheological properties as shear thinning behavior (Stokes et al., 2007), viscoelasticity (Stokes et al., 2007), extensional viscosity (Harward et al., 2010) and normal stress (Stokes et al., 2007). Moreover, the intensity of its properties depends greatly upon the method of stimulation (Stokes et al., 2007).

To discuss about the interest to consider shear thinning behavior in the model, Figure 10 shows the shear rate distribution (calculated by the present model) in the contact between the root of the tongue and the posterior pharyngeal wall for different levels of lubrication by saliva and for mean physiological conditions. When saliva thickness increases, mucosa are more and more close and parallel. At the interface between the food bolus and the saliva, there is a gap of shear stress due to the continuity of shear stress and the difference of viscosity between the two fluids. For the different cases, shear rates vary between 1 and $10^4$ $s^{-1}$, approximately. Stokes et al. (2007) shows that the shear viscosity of saliva vary at maximum between 20 and 1 mPa.s for shear rates comprise between 2 and $5.10^3$ $s^{-1}$. These variations are relatively important; knowing that, the thickness of product varies with the square of the viscosity in the monolayer case. At the light of the present results, the shear thinning behavior of saliva should change quantitatively the model predictions.

Saliva has a highly elastic nature (Stokes et al., 2007) that has to be compared to the time scale of the coating process during swallowing. This time scale is given by the ratio l'/U', where l' is the length of the contact (≈10 mm, Figure 10), is about 20 ms. For saliva, Stokes et al. (2007) reported that the relaxation times of saliva are from 30 ms to 1 s. Being superior to the time scale of the pharyngeal mucosa coating process, viscoelasticity can have an influence on the coating phenomena.

Saliva presents also an extensional viscosity $\mu_E'$ (Harward et al., 2010). According to the results of Harward et al. (2010), the extensional viscosity depends on the strain rate and can reach 120 times the shear viscosity. In the momentum conservation equation, we can demonstrate that the ratio of the stresses due to the extensional viscosity to the shear viscosity is given by $\frac{\mu_E'}{\mu_1'}\left(\frac{h'}{l'}\right)^2$, where h' is the gap between the surfaces (≈100 µm, Figure 10). The value of this ratio is about 0.01 (<<1). We can conclude that extensional effects of saliva should have a slight effect on the coating of mucosa.

The shear of saliva induces normal stress effects (Stokes et al., 2007) that could participate to support the load L' applied by the constrictor muscles. Normal stress $N_1'$ is about 10-100 Pa for shear stresses comprise between 10 and 2000 s$^{-1}$ (Stokes et al., 2007). In the contact between the roots of the tongue and the posterior pharyngeal wall, these effects could generate a load $L_{N1}'$ given by $N_1'.l'$, approximately. An order of magnitude of $L_{N1}'$ is 1 N/m. This value represents only 10% of the load $L'$ applied by the constrictor muscles. We can so conclude that the normal stress effects of saliva must have a moderate effect on the coating phenomena during swallowing.

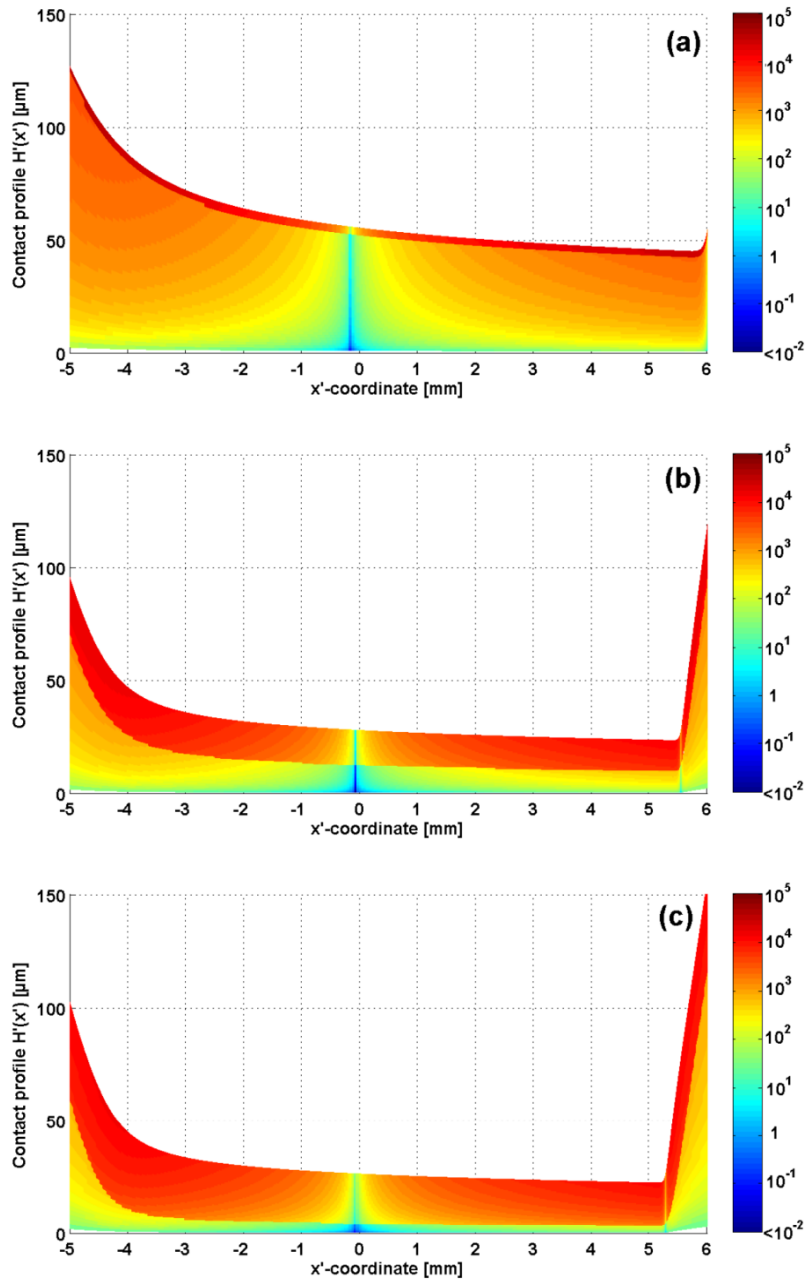

**Figure 10:** Example of shear rate distribution (isovalues of shear rate in 1/s) in the contact for different level of saliva lubrication: $e_1$=2.6 µm and $e_2$=61 µm (a), $e_1$=13 µm and $e_2$=17 µm (b), $e_1$=22 µm and $e_2$=4.2 µm (c). The z'-coordinate 0 correspond to the axis of symmetry (U'=0.5 m/s, E'=20 kP a, $e_m'$=4 mm, R'=4 mm, $\mu_1'$=5 mPa.s, $\mu_2'$=10 mPa.s, L'=10 N/m).

Thus, at the light of the simulations obtained with the present model, we can conclude that the shear thinning behavior and the viscoelasticity of saliva should affect mucosa coating phenomena and would be interesting to study in detail. However, these phenomena could affect the results only quantitatively. In fact, qualitatively, the existence of the two sets of conditions demonstrated in this study is due to obstructions effects by saliva. Moreover, the behavior law of saliva has to be determined on a large scale of shear rates that is difficult to obtain experimentally and the effect of viscoelasticty on lubrication-flows characteristics is a "largely-unresolved problem" (Zhang and Li, 2005).

*5.2. Rheology of food bolus*

A second interesting question is the role of food bolus rheology on coating phenomena. Food bolus can present all kind of rheological properties from liquid to semi-solid food products or chewing solid food. In the present model, we choose to only explore the viscous effects in order to not over-sophisticate the model and to be representative of the experimental conditions of Doyennette et al. (2011) and compare thus the results of these two different approaches.

However, as saliva, it is clear that more complex rheological properties can impact on coating phenomena. For example, biopolymers and hydrocolloids used as thickeners present shear thinning behaviors. Food bolus can also present a yield stress. The yield stress effects and the shear thinning behavior can have a great impact on the coating phenomena because the shear rates generated in the contact vary from 0 to $10^4$ $s^{-1}$. It could be interesting to develop a specific experimental device as in our previous study (de Loubens et al., 2010) with deformable rolls to study the influence of complex rheological properties on coating (as inhomogeneous food bolus for example). To study pharyngeal mucosa coating, modeling stays an interesting approach because it allows us to evaluate physical quantities that are very difficult to measure *in vivo*.

## 6. Conclusion

To conclude, the elastohydrodynamic model of swallowing provides physical explanations as to the role of saliva on the food bolus coating and flavour release. After being successfully compared with *in vivo* experiments, this type of approach is promising for designing food products with specific aroma release kinetics or for adapting food product properties to people who suffer from swallowing disorders. However, the food bolus presents complex behaviours and the development of *in vitro* systems to model swallowing may be of great interest for studying the role of the rheological properties of the food bolus on the pharyngeal mucosa coating and flavour release.

## Acknowledgments

The authors gratefully acknowledge the French National Research Agency (ANR) project SensInMouth for financial support.